\documentclass[3p,times,procedia]{elsarticle}
\usepackage{nupha_ecrc}

\volume{00}

\firstpage{1}

\journalname{Nuclear Physics A}

\runauth{N. Jacazio on behalf of the ALICE collaboration}

\jid{nupha}

\jnltitlelogo{Nuclear Physics A}

\usepackage{amssymb}
\usepackage{graphicx}
\usepackage{calc}
\usepackage{caption}
\usepackage{subcaption}
\usepackage[figuresright]{rotating}

\newcommand{\pT}{\ensuremath{p_{\rm T}}}
\newcommand{\MeVc}{\ensuremath{{\rm MeV/}c}}
\newcommand{\GeVc}{\ensuremath{{\rm GeV/}c}}

\newcommand{\PbPb}{Pb--Pb}
\newcommand{\pPb}{p--Pb}
\newcommand{\pp}{pp}
\newcommand{\sqrtsNN}{\ensuremath{\sqrt{s_{\rm NN}}}}
\newcommand{\sqrts}{\ensuremath{\sqrt{s}}}
\newcommand{\cTeV}{5.02~TeV}
\newcommand{\dTeV}{2.76~TeV}
\newcommand{\sTeV}{7~TeV}

\newcommand{\sqrtssTeV}{\sqrts~=~\sTeV}
\newcommand{\sqrtsNNcTeV}{\sqrtsNN~=~\cTeV}
\newcommand{\sqrtsNNdTeV}{\sqrtsNN~=~\dTeV}
\newcommand{\raa}{\ensuremath{R_{AA}}}
\newlength{\lengtha}
\newlength{\lengthb}
\newcommand{\commonwidth}{\lengtha - \lengthb}%Column width
\begin{document}
  
  \begin{frontmatter}
    
    
    
    %% Instructions from Editor: Please use the following \dochead only in the preprint version (e-print arXiv etc.); 
    %% use empty \dochead{} when submitting to Nuclear Physics A!
    \dochead{XXVIth International Conference on Ultrarelativistic Nucleus-Nucleus Collisions\\ (Quark Matter 2017)}
    % \dochead{}
    %% e.g. \dochead{17th International Conference on Dynamical Processes in Excited States of Solids}
    
    \title{Production of identified and unidentified charged hadrons in \PbPb\ collisions at \sqrtsNN ~5.02~TeV}
    
    
    \author[label1,label2]{Nicol\`o Jacazio}
    
    \address[label1]{Universit\`a di Bologna - Bologna, Italy}
    \address[label2]{INFN, Sezione di Bologna - Bologna, Italy}
    
    \begin{abstract}
      %       In late 2015, the ALICE collaboration recorded data from \PbPb\ collisions at the unprecedented energy of \sqrtsNNcTeV\ as well as reference data from \pp\ collisions at the same energy. 
      In late 2015, the ALICE collaboration recorded data from \PbPb\ collisions at the unprecedented energy of \sqrtsNNcTeV . 
      The transverse-momentum (\pT ) spectra %of unidentified charged hadrons as well as 
      of pions, kaons and protons are presented. 
      The evolution of the particle ratios as a function of collision energy and centrality is discussed. 
      The ratio between \pT -integrated particle yields are measured and compared to different collision energies as well as smaller collision systems.
      For the study of energy loss mechanisms in the QCD medium at high transverse momenta, the nuclear modification factors (\raa ) are computed and compared with results obtained at lower energy.
    \end{abstract}
    
    
  \end{frontmatter}
  
  %   \linenumbers
  
  \section{Introduction}
  \label{sec:Introduction}
  The ultimate goal of heavy-ion physics is the study of the properties of the Quark-Gluon Plasma (QGP), a de-confined and chirally restored state of matter. The measurement of the transverse momentum (\pT ) spectra of (un-)identified particles provides a solid understanding of the collective properties and of the particle production in the fireball created in heavy-ion collisions which is necessary for the correct interpretation of many signatures of QGP creation.
  The latest heavy-ion runs at the LHC, concluded in 2015, gave the possibility to record \PbPb\ collisions at the highest energy ever achieved in the laboratory, allowing the quantitative comparison with lower energy collisions but also with smaller collision systems such as proton-proton (\pp ) and proton-lead (\pPb ).
  The ALICE experiment \cite{ALICEExp, ALICEperf} is particularly well suited to study the production of both identified and unidentified charged particles thanks to its excellent tracking performance coupled with extensive particle identification (PID) capabilities over a wide range of transverse momentum. 
  Particles are identified by combining different techniques allowing their continuous separation over a large transverse momentum interval.
  The Inner Tracking System (ITS) and the Time Projection Chamber (TPC) detectors allow one to identify particles in the lower \pT\ region (starting from 100 \MeVc ) by measuring their specific energy loss.
  The Time Of Flight detector (TOF) which measures the particle velocity and the High Momentum Particle IDentification detector (HMPID) which measures the angle of emission of Cherenkov light are used to extend the PID to higher momentum (up to 6 \GeVc ).
  Finally, the relativistic rise of the energy loss in the TPC gas can be used to further separate different particle species up to 12 \GeVc .
  
  \section{Data analysis and results}
  \label{sec:Results}
\begin{figure}
    \centering
    \begin{subfigure}[b]{.82\textwidth}
      \includegraphics[trim = 2 1 20 15, clip, width=\textwidth]{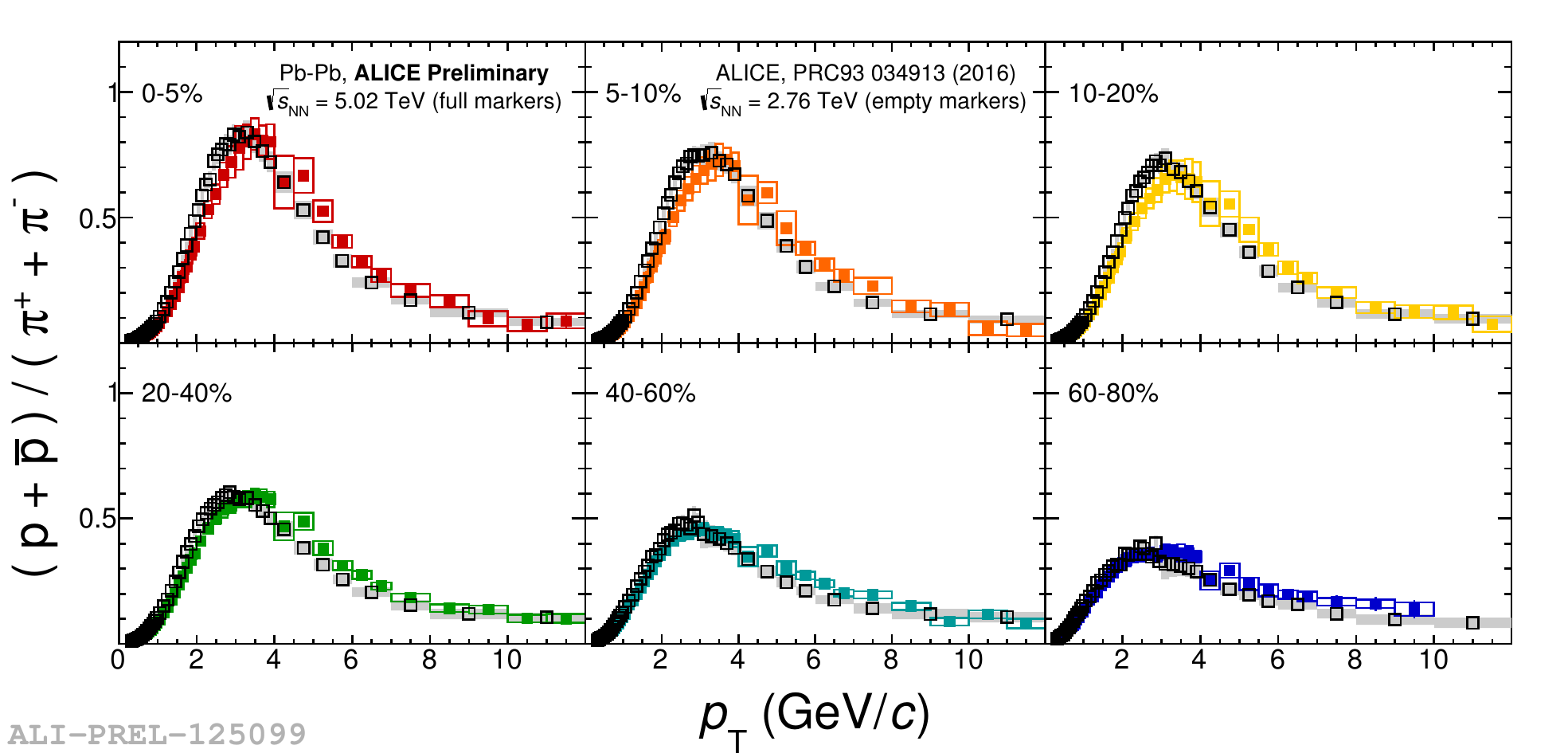}
      \caption{}
      \label{fig:pOverpi}
    \end{subfigure}
    
    \vspace{.2cm}
    
    \begin{subfigure}[b]{.82\textwidth}
      \includegraphics[trim = 2 1 20 15, clip, width=\textwidth]{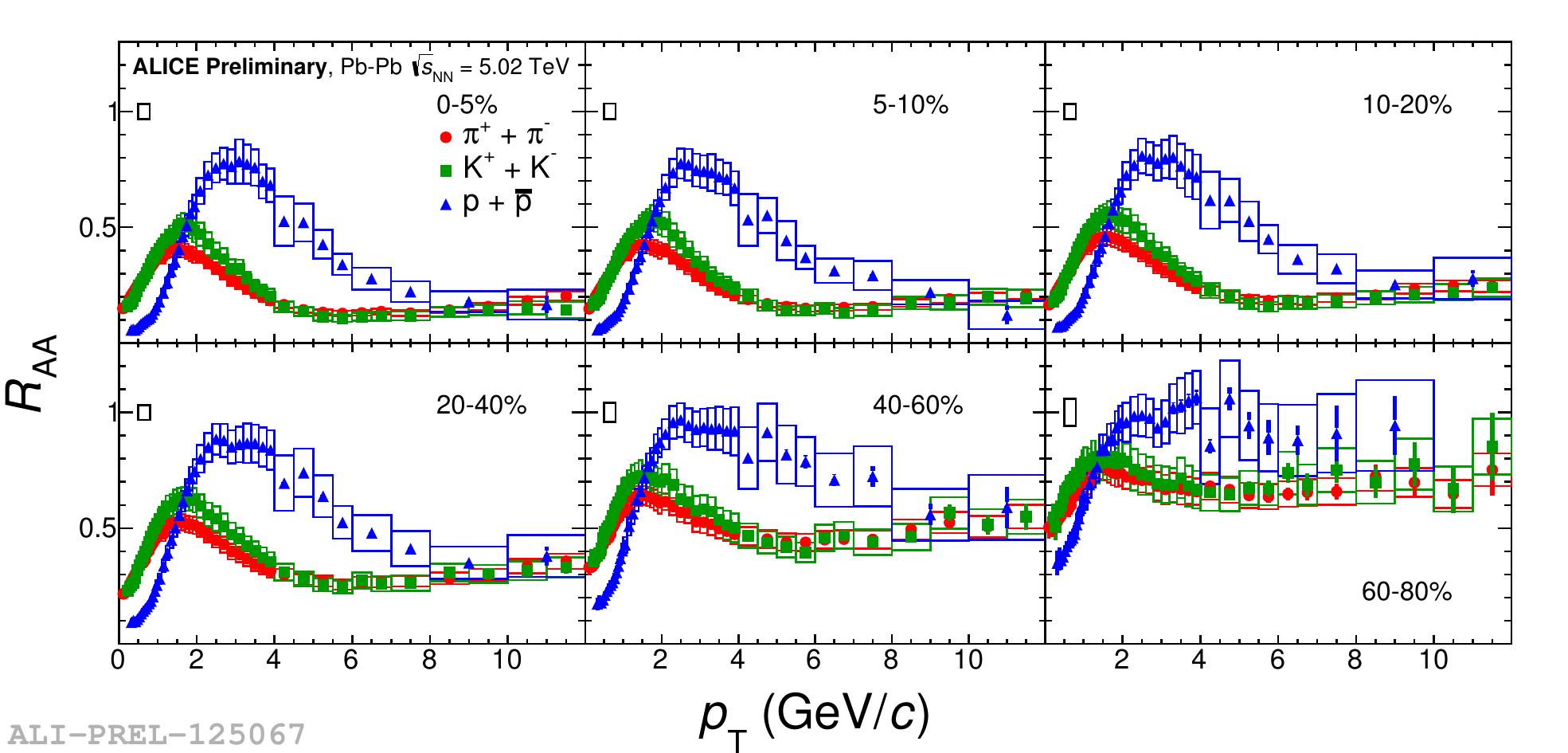}
      \caption{}
      \label{fig:IDRAA}
    \end{subfigure}
    \caption{
      (a) Spectra of identified protons scaled to the ones of pions for different centrality classes, as a function of \pT . 
      Coloured markers show the results at \sqrtsNNcTeV\ while black markers represent the values of \sqrtsNNdTeV\ as reported in \cite{ALICEcent276Raa}.      
      (b) Nuclear modification factor ($R_{AA}$) measured in \PbPb\ collisions at \sqrtsNNcTeV\ for identified pions, kaons and protons as a function of \pT\ for different centrality classes.
    }
    \label{fig:spectrapp}
    \bigskip
    %     \centering
    \begin{subfigure}[b]{\commonwidth}
      \includegraphics[trim = 2 3 49 32, clip, width=\textwidth]{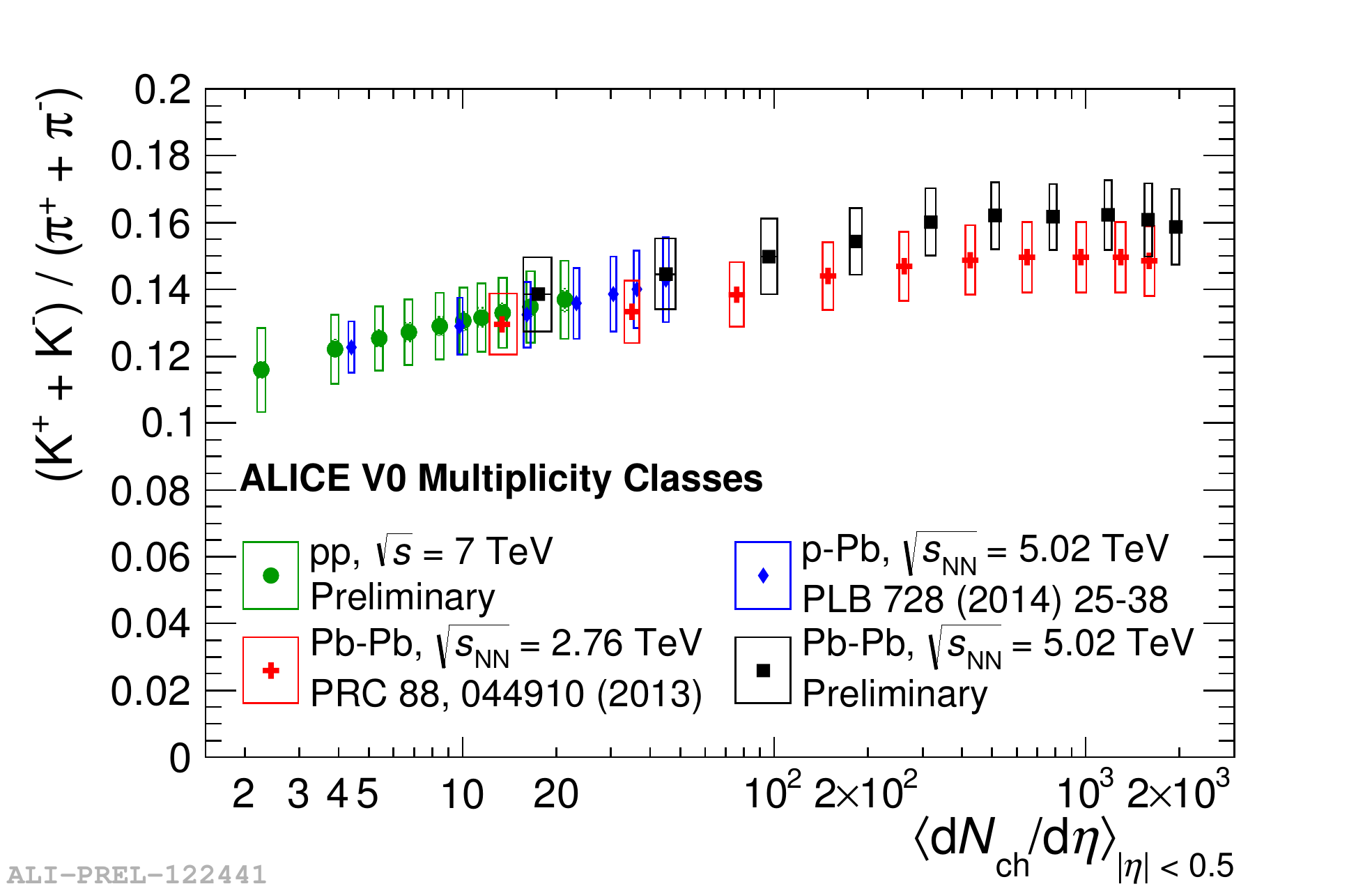}
      \caption{}
      \label{fig:kToPi}
    \end{subfigure}
    %   \hspace{\commonspace}%
    \begin{subfigure}[b]{\commonwidth}
      \includegraphics[trim = 2 3 49 32, clip, width=\textwidth]{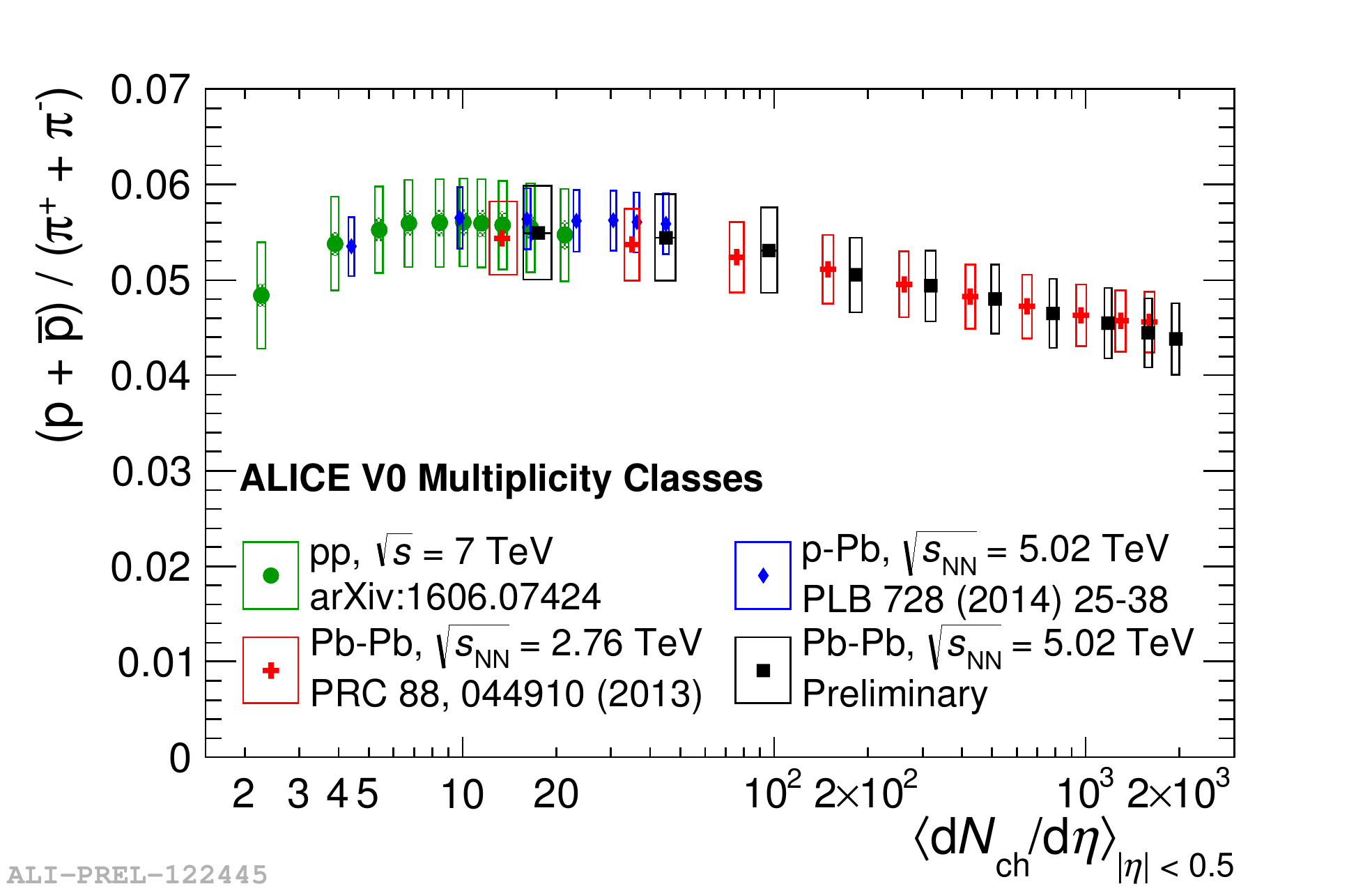}
      \caption{}
      \label{fig:PToPi}
    \end{subfigure}
    \caption{
      \pT -integrated yields of kaons (a) and protons (b) scaled to the ones of pions as a function of the charged particle density for the \pp\ \cite{ALICEpp7TeVvsMult} , \pPb\ and \PbPb\ \cite{ALICEpPb502} (\sqrtsNNdTeV\ \cite{ALICEcent276} and \cTeV ) collisions systems.
    }
    \label{fig:YieldRatios}
  \end{figure}

  We report for the first time on the production of identified $\pi$, K and p measured in \PbPb\ collisions at \sqrtsNNcTeV\ as a function of centrality.
  The data sample was recorded in 2015 with a minimum-bias trigger.
  The total charge collected in the V0 detectors (V0M amplitude), a set of two scintillator hodoscopes located in the pseudorapidity region $2.8 < \eta < 5.1$ (V0A) and $- 3.7 < \eta < - 1.7$ (V0C) and covering the full azimuth was used to determine the centrality of each \PbPb\ collision defined as percentiles of the total hadronic cross section.
  Further details are given in \cite{ALICECent5TeVEstimation, ALICECent276TeVEstimation}.
  Contributions from weak decays of strange particles and from particle knock-out in the material were removed with the data driven approach described in \cite{ALICEcentral276}.
  The systematic uncertainties were estimated by varying the PID techniques and the selection criteria used to define the track sample. 
  The amount of pile-up per event was reduced by selecting runs with low interaction rate and by rejecting events with more than one reconstructed vertex, resulting in a negligible effect.
  The evaluation of the efficiency and acceptance corrections as well was performed using events simulated with the HIJING \cite{HIJING} event generator and embedded into a detailed description of the ALICE detector through which tracks are propagated with the GEANT3 \cite{GEANT3} transport code.
  \begin{figure}
    \centering
    \begin{minipage}{\textwidth}
      \includegraphics[trim = 3 3 28 25, clip, width=.32\textwidth]{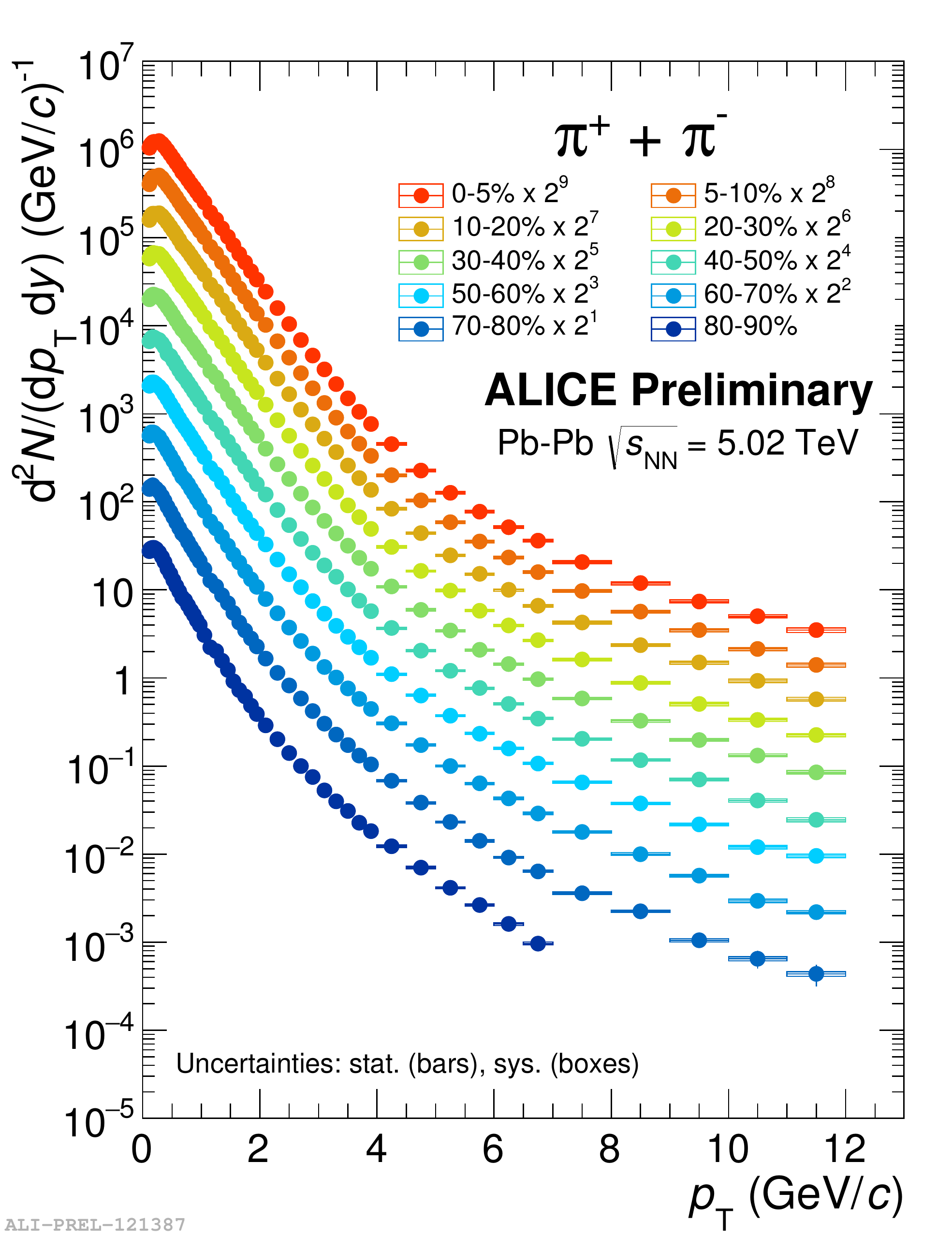}
      \hspace{.01\textwidth}
      \includegraphics[trim = 3 3 28 25, clip, width=.32\textwidth]{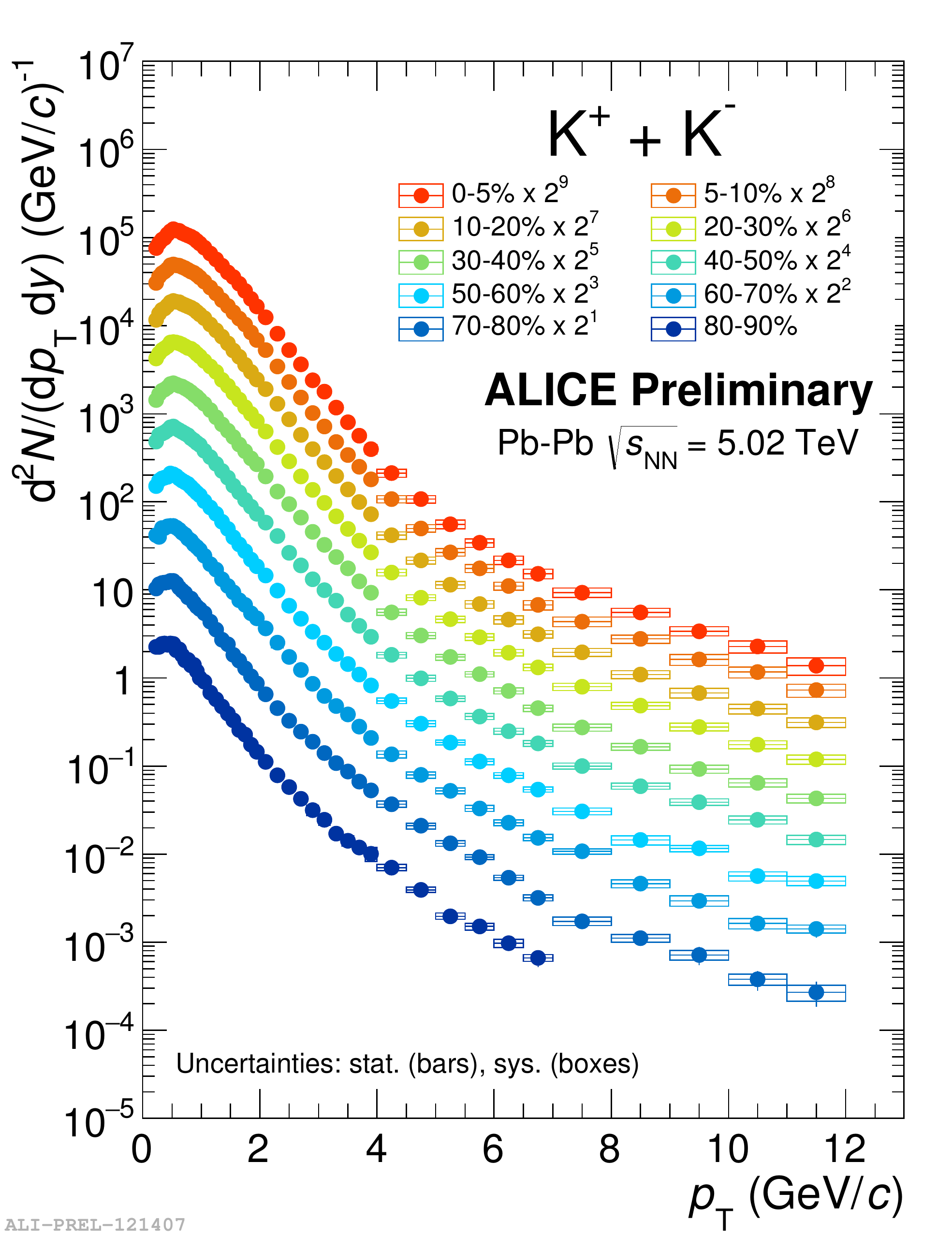}
      \hspace{.01\textwidth}
      \includegraphics[trim = 3 3 28 25, clip, width=.32\textwidth]{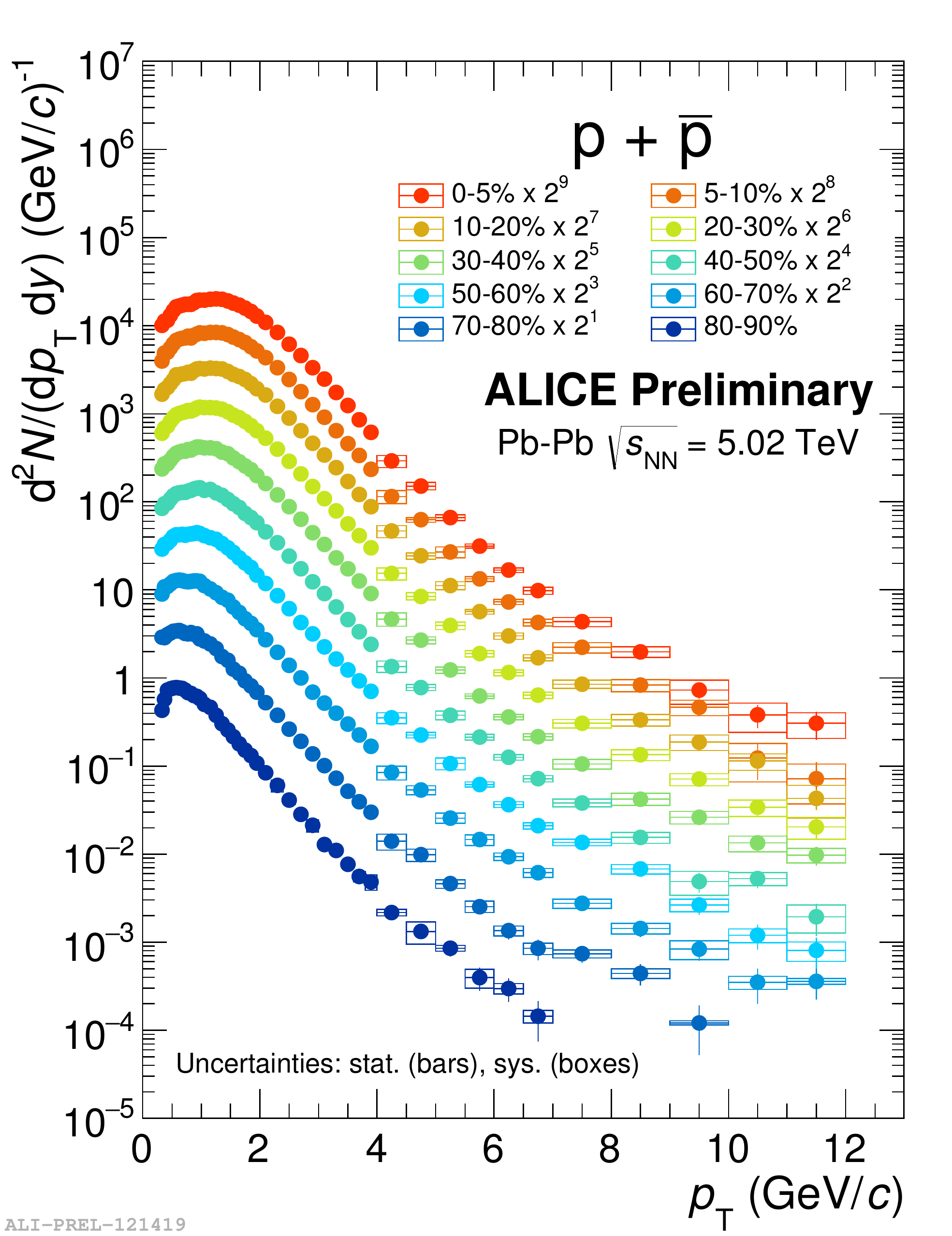}
      \caption{
	\label{fig:IDspectra}
	Spectra of identified charged pions (left panel), kaons (center panel) and protons (right panel) as a function of \pT\ for different centrality classes: warm (cold) colours representing the central (peripheral) collisions, as measured in \PbPb\ collisions at \sqrtsNNcTeV .
      }
    \end{minipage}%
  \end{figure}
  The \pT\ spectra of identified $\pi$, K and p are shown for different centrality classes in Fig. \ref{fig:IDspectra}. 
  Already a direct comparison of the spectral shapes reveals that the spectra become harder with increasing centrality.
  This hardening is found to be mass dependent with protons being more affected than pions. This is consistent with the presence of a radial flow. 
  This effect can be better appreciated in Fig. \ref{fig:pOverpi} where the spectra of protons are divided by the ones of charged pions for each centrality class. 
  From the comparison with the $p/\pi$ ratio at \sqrtsNNdTeV\ shown in Fig. \ref{fig:pOverpi} one can deduce that the effect of the radial flow is larger at higher energies.
  Similar conclusions can be drawn from the analysis of the spectral shape in the framework of the Blast-Wave model \cite{bgbw} which yields a slightly larger transverse expansion velocity for the most central events.

  \newpage
  The nuclear modification factors (\raa ) are shown for pions, kaons and protons as a function of \pT\ for different centrality classes in Fig. \ref{fig:IDRAA}.
  As in the case of the \raa\ measured for unidentified charged particles at \sqrtsNNcTeV\ \cite{ALICEUniVsCent502} the comparison of these results with the ones at \sqrtsNNdTeV\ \cite{ALICEcent276Raa} shows no significant dependence on the collision energy.
  The $R_{AA}$ at large \pT\ $\left( \pT > 8\ \GeVc \right)$ shows the same suppression, independently on the particle species.
  The new results confirm the ones observed at \sqrtsNNdTeV .
  
  The \pT -integrated yields of kaons and protons scaled to the ones of pions for \pp\ \cite{ALICEpp7TeVvsMult} (\sqrtssTeV ), \pPb\ \cite{ALICEpPb502} (\sqrtsNNcTeV ) and \PbPb\ (\sqrtsNNdTeV\ \cite{ALICEcent276},  \sqrtsNNcTeV ) collisions are shown as a function of the density of charged particles at midrapidity in Figs. \ref{fig:kToPi} and \ref{fig:PToPi}, respectively.
  While the comparison among different systems highlights a continuous evolution as a function of the particle multiplicity, the new data confirms the trends observed at lower energy.
  The observation of similar trends at \sqrtsNNdTeV\ and \sqrtsNNcTeV\ is consistent with the expectations from the thermal-statistical model which predicts no significant evolution between the two energies, nevertheless it also shows that the previously observed differences between the data and the model predictions are still present with this new data set at higher energy.
  \section{Conclusions}
  \label{sec:conclusions}
  
  The ALICE Collaboration has presented the results on the production of identified pions, kaons and protons measured as a function of the event centrality in \PbPb\ collisions at \sqrtsNNcTeV .
  A significant evolution of the spectral shape is observed when going from peripheral to central collisions resulting in a blueshift of the spectra.
  The spectra hardening is shown to be mass dependent, being more pronounced for heavier particles, consistently with the picture of the radial flow.
  The comparison with analogous measurement performed in \PbPb\ collisions at \sqrtsNNdTeV\ reveals a slightly larger radial expansion velocity at higher energy.
  This observation is backed up by the analysis of the spectra in the framework of the Blast-Wave model which yields a larger velocity of expansion for the fireball in the most central collisions.
  The ratios of \pT -integrated particle yields (p$/\pi$ and K$/\pi$) measured as a function of the charged particle density show a continuous evolution which seems to be dependent only on the multiplicity and not on the system size.
  Furthermore, the comparison of the new data with \PbPb\ collisions at \sqrtsNNdTeV\ shows no significant energy dependence within the systematic uncertainty.  
  In conclusion, the results presented in these proceedings consolidate the observations performed at lower energy where differences between the data and model predictions, such as relative particle yields, were revealed.

  
  
  %% References
  
  
  \bibliographystyle{elsarticle-num}
  \bibliography{bibliography}
  %   \bibliography{bibliographywTitle}
  
  
  
  
  
  
  % \end{thebibliography}
  
\end{document}